\documentclass[seceq]{jpsj3}
\usepackage{bm}
\usepackage{color}

\newcommand{\rmB}{{\rm B}}

\newcommand{\rmd}{{\rm d}}
\newcommand{\rmtr}{{\rm tr}\,}
\newcommand{\1}{\mbox{1}\hspace{-0.25em}\mbox{l}}
\newcommand{\ky}{{k_y}}

\newcommand{\ind}{{\rm ind}\,}
\newcommand{\rmTr}{{\rm Tr}\,}

\title{
Bulk-Edge Correspondence for Chern Topological Phases:\\
A Viewpoint from a Generalized Index Theorem
}

\author{
\name{
Takahiro \surname{Fukui}$^1$, Ken \surname{Shiozaki}$^2$, 
Takanori \surname{Fujiwara}$^1$, and Satoshi \surname{Fujimoto}$^2$
} 
}
\inst{
\address{
$^1$Department of Physics, Ibaraki University, Mito 310-8512, Japan\\
$^2$Department of Physics, Kyoto University, Kyoto 606-8502, Japan
} 
}
\abst{
We explore the bulk-edge correspondence for 
topological insulators (superconductors) without time-reversal symmetry
from the viewpoint of the 
index theorem for open spaces.  We assume generic Hamiltonians not only with a linear dispersion 
but also with higher order derivatives arising from generic band structures. 
Using a generalized index theorem valid for such systems,
we show the equivalence between the spectral flow 
of the edge states and the Chern numbers specifying the bulk systems.
}

\kword{bulk-edge correspondence, topological insulators, index theorem, spectral flow, Chern number,
spectral asymmetry
}

\begin{document}
\maketitle

\section{Introduction}

When a bulk gapped system has a topologically nontrivial ground state,
gapless edge states appear if the system has boundaries. 
This bulk-edge correspondence has become increasingly important since 
the recent discovery of various types of topological insulators and superconductors.\cite{HasanKane,QiZhang,SRFL08,Kitaev08,Wen12}
For quantum Hall systems, the relationship between the Chern number of Landau 
levels\cite{TKNN,Kohmoto85} and the quantized Hall conductance as a result of 
the edge current\cite{Halperin82}
was elucidated by Hatsugai \cite{Hatsugai93} for a tight-binding model.
Volovik\cite{Volovik03} and recently Essin and Gurarie\cite{EssGur11} 
formulated the correspondence by using the Green functions
for more generic systems with a domain wall separating topologically distinct systems.
An adiabatic approach was also developed by Sato {\it et al.} for topological superconductors.\cite{STYY11}

More generically, defects such as kinks, vortices, and monopoles can be regarded as a type of boundary,
and indeed they allow midgap states. 
In particular, zero-energy states are known to
reflect the topological property of the corresponding bulk systems.
These zero-energy states were investigated by index theorems that show the equivalence between
the analytical index concerning the zero-energy states as solutions of differential equations
and the topological index associated with the topological configurations of background fields
included in the operators (Hamiltonians).
Therefore, it is natural to utilize the index theorems to investigate 
the problem of edge states associated with the conventional boundary mentioned above.

In this paper, we develop a method of investigating the bulk-edge correspondence 
for class A insulators and class D superconductors
by using the index theorems formulated by Callias,\cite{Callias78} 
Weinberg,\cite{Weinberg81} and Niemi and Semenoff\cite{NieSem86}, and recently applied to 
the Majorana states of topological superconductors \cite{FukFuj10,IFF12,SFF12},
which are valid for open infinite systems in arbitrary dimensions.
Edge states are usually described as follows:
Suppose that 
a translationally invariant boundary lies along the $y$-axis for simplicity. Then,
we have a one-dimensional Hamiltonian with a parameter $k_y$. Edge states are defined by the spectral flow
of this Hamiltonian as a function of $k_y$. 
To utilize the index theorems for such systems, one often regards the parameter 
of the Hamiltonian as a part of background fields. 
However, since the parameter in the present case, $k_y$, does not have upper and lower bounds,
it is difficult to calculate the topological index in a standard manner such as by derivative expansions.
Furthermore, 
the index theorems \cite{Callias78,Weinberg81,NieSem86} studied thus far were restricted
to the Dirac operator (Hamiltonian) only with
the first-order derivatives, whereas in condensed matter physics, we often encounter 
systems with higher derivatives arising from general band structures. 
The purpose of this paper is
to develop a generalized index theorem applicable to such generic Hamiltonians, and to present
several techniques invented for 
its application to the bulk-edge problem.

This paper is organized as follows:
In \S \ref{s:BulEdg}, we present the setup of the bulk-edge correspondence for a generic Hamiltonian.
We first introduce a bulk Hamiltonian whose topological properties are controlled by one parameter.
We next consider a Hamiltonian with a domain wall that separates topologically distinct systems,
by making the parameter coordinate-dependent. This Hamiltonian is referred to as a domain wall Hamiltonian.
For this Hamiltonian, edge states are expected at the domain wall. 
To characterize the edge states, we define the spectral flow of these states.
In \S \ref{s:CheNum}, we present the formulas for the Chern number characterizing the bulk systems.
In \S \ref{s:IndEdg}, we first extend the domain wall Hamiltonian, and 
show that the edge states can be regarded as the zero-energy states of this extended Hamiltonian.
This implies that the edge states are characterized by the index of the extended Hamiltonian.
In \S \ref{s:IndThe}, we first derive a generalized index theorem and next 
apply it to the extended Hamiltonian. We show that 
the index is given by the Chern numbers, and, together with the fact that
the spectral flow of the edge states is equal to the index of the extended Hamiltonian, 
we thus establish the bulk-edge correspondence.

\section{Setup for Bulk-Edge Correspondence}
\label{s:BulEdg}

We investigate a generic bulk Hamiltonian
${\cal H}^\rmB(-i\partial_x,-i\partial_y)$
depending on a constant parameter $m$ 
controlling the topological properties of the bulk system.
Without loss of generality, we assume that the gap is open at the zero energy and that
negative energy states are occupied. 
We also assume translational invariance, which enables us
to represent the Hamiltonian in the momentum space such that
\begin{alignat}1
{\cal H}^\rmB(k_x,k_y)\equiv{\cal H}^\rmB(k),
\label{BulHamMom}
\end{alignat}
where $k$ stands for the set of two momenta, $k_x$ and $k_y$.
Then, the occupied bands give rise to a nontrivial Chern (TKNN) number $c(m)$,
which shall be discussed in the next section.  
Correspondingly, we introduce a model with a domain wall at $x=0$,
which separates two different topological states, i.e.,  
\begin{alignat}1
\lim_{x\rightarrow\pm\infty} m(x)=m_\pm .
\label{InfMas}
\end{alignat}
Let us denote such a Hamiltonian as
${\cal H}(-i\partial_x,-i\partial_y,x)$, 
where the $x$ dependence implies that the parameter $m$ is a function of $x$.
We assume that the Hamiltonian is still translationally invariant along the $y$ direction
in the presence of the domain wall,
and hence, the Fourier transformation leads to the Hamiltonian 
\begin{alignat}1
{\cal H}(-i\partial_x,\ky,x) .
\label{DomWalHam}
\end{alignat}
This is a one-parameter family of one-dimensional Hamiltonians and its spectrum as a function of the 
parameter $k_y$ includes edge states localized around the domain wall.
The eigenvalue equation is 
\begin{alignat}1
{\cal H}(-i\partial_x,\ky,x)\phi_j(x,\ky)=\varepsilon_j(\ky)\phi_j(x,\ky).
\label{EigValEquEdg}
\end{alignat}
Although the bulk system has a gap at the zero energy, 
eq. (\ref{EigValEquEdg}) admits several states that cross the zero energy. 
These states are referred to as edge states. More precisely, 
the edge states are those with the spectrum 
$\varepsilon_j(-\Lambda_{j{\rm L}})\varepsilon_j(\Lambda_{j{\rm U}})<0$, 
where $-\Lambda_{j{\rm L}}$ and $\Lambda_{j{\rm U}}$ are the lower and upper bounds of $k_y$ within which
$\varepsilon_{j}(k_y)$ is defined as a continuous function.
\cite{footnote2}
Assume that the model has $N_+$ edge states with $\varepsilon_j(\Lambda_{j{\rm U}})>0$,
and $N_-$ states with $\varepsilon_j(\Lambda_{j{\rm U}})<0$.
The spectral flow  $\nu$ is defined by
\begin{alignat}1
\nu=N_+-N_-.
\label{DefEta}
\end{alignat}
The relationship between the Chern number $c(m_\pm)$ of the bulk system (\ref{BulHamMom})
and the spectral flow $\nu$ of the domain wall model (\ref{DomWalHam}) 
is the generic bulk-edge correspondence investigated in this paper.

In what follows, we suppose a Dirac Hamiltonian or 
a more generic Hamiltonian of a Dirac-like matrix structure with higher-derivative terms.
For later convenience, we introduce the following notation separating the derivative terms from the others,
\begin{alignat}1
{\cal H}(-i\partial_x,k_y,x)&=\sum_{n\ge1}\gamma_x^{(n)}(-i\partial_x)^n+\delta{\cal H}(k_y,x).
\label{HamSep}
\end{alignat}
We have assumed that $m(x)$ is not coupled with the derivative terms.
The coefficients of the derivative terms $\gamma_x^{(n)}$ may be a function of $k_y$.
The $k_y$ dependence of $\gamma_x^{(n)}$ does not change the following argument.

A nontrivial example of the bulk Hamiltonian is \cite{BHZ:06,RHV:10,QHZ10,LSYNS10,SSL11,Qi:11,FHNQ:11} 
%
${\cal H}^\rmB=-i\sigma_j\partial_j+\sigma_3\left(m+b\partial^2\right)$,
where $j=1,2$, $(x_1=x, x_2=y)$ and $\partial^2\equiv \partial_x^2+\partial_y^2$.
Its momentum representation is
\begin{alignat}1
{\cal H}^\rmB(k)=\sigma_jk_j+\sigma_3(m-bk^2),
\label{BulHam}
\end{alignat}
where $k^2\equiv k_x^2+k_y^2$.
The corresponding domain wall model is therefore
\begin{alignat}1
{\cal H}(-i\partial_x,k_y,x)=-i\sigma_1\partial_x+\sigma_2\ky
+\sigma_3\left[m(x)-bk_y^2+b\partial_x^2\right].
\label{ConDomWalMod}
\end{alignat}
Comparing with eq. (\ref{HamSep}), we have
\begin{alignat}1
&\gamma_x^{(1)}=\sigma_1,\quad \gamma_x^{(2)}=-b\sigma_3,
\nonumber\\
&\delta{\cal H}(k_y,x)=\sigma_2k_y+\sigma_3\left[m(x)-bk_y^2\right].
\label{ConDomWalModEac}
\end{alignat}
As shown in the next section, the bulk Hamiltonian (\ref{BulHam}) has a nontrivial Chern number.

\section{Chern Number of the Bulk System}
\label{s:CheNum}

In this section,
we summarize notational conventions for the Chern number characterizing the bulk system.\cite{WilZee84,Hatsugai04}
Let $\psi_\alpha(k)$ be the normalized eigenstate of ${\cal H}^\rmB(k)$ with the eigenvalue
$\varepsilon_\alpha(k)$,
\begin{alignat}1
{\cal H}^\rmB(k)\psi_\alpha(k)=\varepsilon_\alpha(k)\psi_\alpha(k),
\label{BulEigValEqu}
\end{alignat}
where $\alpha$ specifies the band of the spectrum, 
and let $\psi(k)=(\psi_1,\psi_2,\cdots)$ 
be the set of wave functions with negative energies, $\varepsilon_\alpha(k)<0$.
Then, 
\begin{alignat}1
P=\psi\psi^\dagger 
\label{ProOpe}
\end{alignat}
serves as a projection operator to the negative energy bands.
The Berry connection one-form $A$ and the curvature two-form $F$ are defined by
\begin{alignat}1
&A=\psi^\dagger \rmd \psi ,
\nonumber\\
&F=\rmd A+A^2=\rmd\psi^\dagger(1-P)\rmd\psi ,
\label{DefBer}
\end{alignat}
where $\rmd$ is the exterior derivative with respect to the momentum $k_i$.
The Chern number is defined by
\begin{alignat}1
c&=\frac{i}{2\pi}\int\rmtr F
\nonumber\\&
=\frac{i}{2\pi}\epsilon_{ij}\sum_{\varepsilon_\alpha<0<\varepsilon_\beta}\int \rmd^2k\,
(\partial_{k_i}\psi_\alpha^\dagger)\psi_\beta\psi_\beta^\dagger(\partial_{k_j}\psi_\alpha) ,
\label{DefCheNum}
\end{alignat}
where in the second line, we have used the fact that
$1-P$ in eq. (\ref{DefBer}) is the projection operator to the positive energy states.

The following relation is also useful:
\begin{alignat}1
P(\rmd P)^2=\psi F\psi^\dagger.
\end{alignat}
It follows that $\rmtr F=\rmtr P(\rmd P)^2$.
In the case of the model in eq. (\ref{BulHam}), this formula is useful, since
the projection operator (\ref{ProOpe}) can be expressed alternatively as
$P=\frac{1}{2}\left(1-{\cal H}^\rmB/R\right) $, 
where $R(k)$ is defined by $({\cal H}^\rmB)^2=R^2\1$ ($R>0$) and given concretely by 
$R(k)=\left[k^2+(m-bk^2)^2\right]^{1/2}$.
The Chern number is thus given by
\begin{alignat}1
\nonumber\\
c(m)&=\frac{1}{4\pi}\int\rmd^2k\frac{m+bk^2}{\left[k^2+(m-bk^2)^2\right]^{3/2}}
\nonumber\\
&=\frac{1}{2}\left[{\rm sgn}(m)+{\rm sgn} (b)\right] .
\end{alignat}
Therefore, if $m_+$ and $m_-$ in eq. (\ref{InfMas}) have opposite signs,
a domain wall appears in model (\ref{ConDomWalMod}) that separates 
two distinct systems with trivial and nontrivial Chern numbers.

\section{Index of the Edge States}
\label{s:IndEdg}

Thus far, we have defined the spectral flow (\ref{DefEta}) for the domain wall system as well as 
the Chern number (\ref{DefCheNum}) for the gapped ground state of the bulk system.
To prove the correspondence between them, we introduce an extended Hamiltonian,
\begin{alignat}1
\widetilde{\cal H} &\equiv-i v\tau_1\partial_\ky +\tau_2{\cal H}(-i\partial_x,\ky,x) ,
\label{ExtHam}
\end{alignat}
where $v~(>0)$ is a constant parameter. 
Although ${\cal H}$ is a one-parameter family of one-dimensional Hamiltonians, 
$\widetilde{\cal H}$ is now a two-dimensional model. This technique was used in many references
dealing with index theorems.
\cite{APS76,Witten82,AGPM85,Forte87} 
It should be noted that this Hamiltonian has a generalized chiral symmetry,  
$\{\widetilde{\cal H},\tau_3\}=0$.
This enables us to regard the spectral flow of ${\cal H}$
as an index of the extended Hamiltonian $\widetilde{\cal H}$.

To be specific,
let us consider the zero energy states of this extended Hamiltonian,
\begin{alignat}1
\widetilde{\cal H}\Psi_0(\ky,x)=0 .
\end{alignat}
Owing to the chiral symmetry of $\widetilde{\cal H}$, these states can also be eigenstates of $\tau_3$,
and hence, we assume the following product wave functions:
\begin{alignat}1
\Psi_{0,j}=
\left(\begin{array}{c}{\varphi}_{0,j}^+\\ 0\end{array}\right)\phi_j, \quad
\left(\begin{array}{c}0\\{\varphi}_{0,j}^-\end{array}\right)\phi_j,
\end{alignat}
where $\phi_j$ is the wave function in eq. (\ref{EigValEquEdg}), and 
the former and latter in the above equation have $\tau^3=1$ and $\tau^3=-1$,  respectively. 
By applying an adiabatic deformation preserving topological invariants, we can choose 
$\phi_j$ independently of $k_y$.
Then, it follows that
\begin{alignat}1
\left(v\partial_\ky\mp\varepsilon_j\right)\varphi_{0,j}^\pm=0.
\end{alignat}
The solution of this equation is given by 
\begin{alignat}1
\varphi_{0,j}^\pm(\ky)\propto e^{\pm\frac{1}{v}\int^\ky\rmd\ky'\varepsilon_j(\ky')}.
\end{alignat}
This tells us that for any positive value of $v$, $\psi_0^+$ ($\psi_0^-$)
is normalizable if and only if $\varepsilon_j(-\Lambda_{j{\rm L}})>0>\varepsilon_j(\Lambda_{j{\rm U}})$ 
($\varepsilon_j(-\Lambda_{j{\rm L}})<0<\varepsilon_j(\Lambda_{j{\rm U}})$). 
Therefore, we have
\begin{alignat}1
-\ind \widetilde{\cal H}=\nu .
\label{IndEta}
\end{alignat}
The above argument clearly shows that the index is independent of $v$.
This point plays a crucial role below.

\section{Index Theorem for the Extended Hamiltonian}
\label{s:IndThe}

In the previous section, we introduced the extended model (\ref{ExtHam}) whose 
zero energy states are given by the edge states of the domain wall system.
We next apply the  index theorem, which makes it possible to express the
index as a topological invariant of the Hamiltonian.

\subsection{Index theorem}

We will first generalize the index theorem for the present system 
including higher-order derivatives generically.
Let us denote the extended Hamiltonian (\ref{ExtHam}) simply as
\begin{alignat}1
\widetilde{\cal H}&=-i\gamma_j\partial_j+\delta\widetilde{\cal H},
\label{ReDefExtHam}
\end{alignat} 
where $j=1,2$ and $\partial_j\equiv\partial_{t_j}$ with $(t_1,t_2)=(k_y,x)$.
The $\gamma$ ``matrices'' and $\delta\widetilde{\cal H}$ are defined by
\begin{alignat}1
&
\gamma_1=v\tau_1,\quad \gamma_2=\tau_2\sum_{n\ge1}\gamma_x^{(n)}(-i\partial_2)^{n-1} ,
\nonumber\\ 
&\delta\widetilde{\cal H}=\tau_2\delta{\cal H}(k_y,x) .
\end{alignat}
The chiral symmetry of the Hamiltonian is denoted by $\{\widetilde{\cal H},\tau_3\}=0$.
The index of $\widetilde{\cal H}$ in the previous section can be defined alternatively by
\begin{alignat}1
\ind \widetilde{\cal H}=\lim_{M\rightarrow0}\rmTr \tau_3\frac{iM}{iM-\widetilde{\cal H}},
\label{TopInd}
\end{alignat}
where $\rmTr$ stands for integration over $t~(=k_y,x)$ as well as the trace 
in the $\tau$-space and the internal band structure.
As has been shown in many references,\cite{Callias78,Weinberg81,NieSem86,FukFuj10,IFF12} 
the above index can be expressed in terms of the divergence of 
a chiral current for the Dirac Hamiltonian. 
In what follows, we will show that even if the Hamiltonian includes higher derivatives,
we can define a similar current. 
In the derivation of the index theorem, the formal identity \cite{NieSem86}
\begin{alignat}1
&\lim_{s\rightarrow t}
\rmtr\tau_3\frac{iM}{iM-\widetilde{\cal H}}\delta^2(t-s)
=\lim_{s\rightarrow t}\rmtr \tau_3\delta^2(t-s)
+\lim_{s\rightarrow t}\rmtr\tau_3\frac{\widetilde{\cal H}}{iM-\widetilde{\cal H}}\delta^2(t-s) ,
\label{ForIde}
\end{alignat}
plays a crucial role.
We show in Appendix \ref{s:DivCur} that the last term can be written as 
the divergence of a current 
\begin{alignat}1
\lim_{s\rightarrow t}\rmtr\tau_3\frac{\widetilde{\cal H}}{iM-\widetilde{\cal H}}\delta^2(t-s) =
-\frac{1}{2}\partial_jJ_j,
\label{RewDivCur}
\end{alignat}
where the current is defined by
\begin{alignat}1
&J_j(t)=
\lim_{s\rightarrow t}\rmtr\tau_3\Gamma_j
\frac{i}{iM-\widetilde{\cal H}}\delta^2(t-s) ,
\label{DefCur}
\end{alignat}
with
\begin{alignat}1
\Gamma_1&=\gamma_1=v\tau_1,
\nonumber\\
\Gamma_2&=\tau_2\sum_{n\ge1}\gamma_x^{(n)}p_n(-i\partial_2^t,-i\partial_2^s).
\label{DefExtGam}
\end{alignat}
Note that $\Gamma_2$ includes the effect of the higher derivatives through a 
polynomial $p_n$ defined by
\begin{alignat}1
p_n(a,b)\equiv \sum_{0\le i\le n-1}(-)^{i}a^{n-i-1}b^i.
\label{PolDef}
\end{alignat}
If the model has only the first-order derivative,  $p_1(a,b)=1$ implies that $\Gamma_2$ becomes 
the usual matrix. The reason why we have introduced the polynomial (\ref{PolDef}) is   
to use the following identity:
\begin{alignat}1
p_n(a,b)(a+b)=a^{n}-(-b)^n .
\label{PolIde}
\end{alignat}
This equation is required to derive eq. (\ref{RewDivCur}),
as shown in Appendix \ref{s:DivCur}. 
The index (\ref{TopInd}) can thus be rewritten as
\begin{alignat}1
{\rm ind}\,\widetilde{\cal H}=
\lim_{\mu\rightarrow\infty}\rmTr\tau_3f(\widetilde{\cal H}^2/\mu^2)
-\frac{1}{2}\lim_{M\rightarrow0}
\rmTr\partial_{j}J_j ,
\label{IndFor}
\end{alignat}
where in the r.h.s., the first term is associated with the chiral anomaly 
when ${\cal H}$ is a Dirac Hamiltonian only with linear kinetic terms.
To treat generic systems with or without higher derivative terms in a unified manner,    
we have introduced an ultraviolet regulator $f(x)$, which is a smooth function satisfying
$f(0)=1$ and $f(\infty)=0$.

In what follows, we shall calculate each term in eq. (\ref{IndFor})
separately. In the conventional index theorem,
the derivative expansion is applicable, since the background fields are assumed to be constant
at infinities. In the present Hamiltonian (\ref{ReDefExtHam}), 
the parameter $m$ indeed satisfies this condition owing to eq. (\ref{InfMas}). 
However, $\delta{\cal H}$ diverges in the limit $|k_y|\rightarrow\infty$, as seen from 
eq. (\ref{ConDomWalModEac}).
This implies that we cannot resort to 
this simple technique of derivative expansion. 
Nevertheless, a crucial role is played by the parameter $v$ introduced in eq. (\ref{ExtHam}) instead. 
As we have noted, the index (\ref{IndEta}) holds for any value of $v~(>0)$. 
This makes it possible to expand eq. (\ref{IndFor}) with respect to this parameter.

\subsubsection{Calculation of the $\tau_3$ term}

Let us first consider the $\tau_3$ term in eq. (\ref{IndFor}).
As a regulator function, we choose the standard function $f(x)=e^{-x}$.
To calculate this term in the limit $v\rightarrow 0$, we explicitly write the integration 
over $\ky$ such that
\begin{alignat}1
\rmTr\tau_3e^{-\widetilde{\cal H}^2/\mu^2}
&=\int\rmd\ky\int_{-\infty}^\infty\frac{\rmd\omega}{2\pi}e^{-i\omega\ky}
\rmTr_x\tau_3e^{-\widetilde{\cal H}^2/\mu^2}e^{i\omega\ky},
\label{Ano1}
\end{alignat}
where $\rmTr_x$ stands for the trace $\rmTr$ except for the integration over $k_y$,
and we have used the Fourier representation of the $\delta$ function in eq. (\ref{ForIde}),
$\delta(t_1-s_1)\equiv\delta(k_y-k_y')=\int\frac{\rmd\omega}{2\pi} e^{i(k_y-k_y')\omega}$, and
taken the limit $k_y'\rightarrow k_y$.
Using 
\begin{alignat}1
e^{-i\omega\ky}\widetilde{\cal H}^2e^{i\omega\ky}
=v^2(\omega-i\partial_\ky)^2+v\tau_3\partial_{\ky}{\cal H}+{\cal H}^2,
\end{alignat}
and making the scale transformation $\omega\rightarrow\omega/v$, we can obtain the 
following expansion with respect to $v$:
\begin{alignat}1
&e^{-i\omega\ky}e^{-\widetilde{\cal H}^2/\mu^2}e^{i\omega\ky}
=\left[
1-\frac{v}{\mu^2}(\tau_3\partial_\ky{\cal H}-2i\omega\partial_\ky)
\right]e^{-(\omega^2+{\cal H}^2)/\mu^2}
+O(v^2) .
\end{alignat}
Thus, we obtain the expansion of the $\tau_3$ term with respect to $v$ such that
\begin{alignat}1
&\rmTr\tau_3e^{-\widetilde{\cal H}^2/\mu^2}
=\int\rmd\ky\int_{-\infty}^\infty\frac{\rmd\omega}{2\pi}\frac{1}{v}\rmTr_x\tau_3
\left[
1-\frac{v}{\mu^2}(\tau_3\partial_\ky{\cal H}-2i\omega\partial_\ky)
\right]
e^{-(\omega^2+{\cal H}^2)/\mu^2}
+O(v^2).
\end{alignat}
In the above expansion, 
terms without $\tau_3$ vanish by the trace over $\tau$. Therefore, in the limit $v\rightarrow0$, we have
\begin{alignat}1
\rmTr\tau_3e^{-\widetilde{\cal H}^2/\mu^2}
&=-2\int\rmd\ky\int_{-\infty}^\infty\frac{\rmd\omega}{2\pi}\frac{1}{\mu^2}\rmTr_x
\frac{\partial{\cal H}}{\partial\ky}
e^{-(\omega^2+{\cal H}^2)/\mu^2},
\end{alignat}
where the factor of 2 originates from the trace over $\tau$, implying that $\rmTr_x$ here does not include
this space any longer. The integration over $\omega$ is carried out to give
\begin{alignat}1
\lim_{\mu\rightarrow\infty}\rmTr\tau_3e^{-\widetilde{\cal H}^2/\mu^2}
=-\frac{1}{\sqrt{\pi}}\lim_{\mu\rightarrow\infty}\int\rmd\ky
\rmTr_x\frac{1}{\mu}
\frac{\partial{\cal H}}{\partial\ky}
e^{-{\cal H}^2/\mu^2}.
\label{AnoFin}
\end{alignat}
As we shall show, this term cancels out the next term below and does not contribute to the 
final index.

\subsubsection{Calculation of the $\rmTr\partial_1J_1$ term}

Let us first compute $J_1(k_y,x)$ for any values of $k_y$ and $x$. 
To this end, we introduce the notation $\Omega=(\omega,k_x)$ as conjugate ``momenta'' to $t=(k_y,x)$. 
Then, by using the Fourier representation of the two-dimensional $\delta$ function $\delta^2(t-s)$
in (\ref{DefCur}), we have
\begin{alignat}1
J_1(\ky,x)
&=\int\frac{\rmd^2\Omega}{(2\pi)^2}e^{-i\Omega t}
\rmtr\tau_3\Gamma_1\frac{i}{iM-\widetilde{\cal H}}e^{i\Omega t}
\nonumber\\
&=
-\int\frac{\rmd^2\Omega}{(2\pi)^2}\rmtr v\tau_3
\frac{1}{\tau_1M+iv\omega+v\partial_\ky-\tau_3{\cal H}(k_x-i\partial_x,k_y,x)},
\label{Tsuika}
\end{alignat}
where the minus sign in the last line originates from $\tau_1\tau_3\tau_1=-\tau_3$.
Rescaling $\omega\rightarrow \omega/v$ and expanding with respect to $v$, we see that 
eq. (\ref{Tsuika}) becomes
\begin{alignat}1
J_1(\ky,x)
&=-\int\frac{\rmd^2\Omega}{(2\pi)^2}\rmtr \tau_3
\frac{1}{i\omega+\tau_1M-\tau_3{\cal H}}+O(v)
\nonumber\\
&\rightarrow
2\int\frac{\rmd^2\Omega}{(2\pi)^2}\rmtr
\frac{{\cal H}}{\omega^2+{\cal H}^2},
\end{alignat}
where in the second line, the prefactor of 2 originates from the trace over the $\tau$ space, and
the limit $M\rightarrow0$ has been taken. 
At this stage, we arrive at 
\begin{alignat}1
\lim_{M\rightarrow0}\int\rmd xJ_1(\ky,x)
&=\int_{-\infty}^\infty\frac{\rmd\omega}{\pi}\rmTr_{\!x}
\frac{{\cal H}}{\omega^2+{\cal H}^2}
\nonumber\\
&\equiv \eta(\ky) .
\label{J2Ter}
\end{alignat}
Here, the function $\eta$ is a formal expression of the spectral asymmetry, and
a certain suitable regularization is required.
We adopt \cite{AGPM85,footnote1}
\begin{alignat}1
\eta_s(\ky)=\frac{2}{\Gamma((s+1)/2)}\int_0^\infty\rmd uu^s\rmTr_x{\cal H}e^{-u^2{\cal H}^2}.
\end{alignat} 
In the limit $s\rightarrow0$, $\eta_s\rightarrow \eta_0\equiv\eta$ as calculated above.\cite{footnote1}
It then follows that\cite{AGPM85}
\begin{alignat}1
\frac{\rmd\eta_s}{\rmd\ky}
=\frac{2}{\Gamma((s+1)/2)}\int_0^\infty\rmd uu^s
\partial_u\left(\rmTr_xu\frac{\rmd{\cal H}}{\rmd\ky}
e^{-u^2{\cal H}^2}\right).
\end{alignat}
Since this should be well-defined, we can safely take the limit $s\rightarrow0$, and hence, we obtain
\begin{alignat}1
\frac{\rmd\eta}{\rmd\ky}
&=\frac{2}{\sqrt{\pi}}\int_0^\infty\rmd u\partial_u
\left(\rmTr_xu\frac{\rmd{\cal H}}{\rmd\ky}e^{-u^2{\cal H}^2}\right)
\nonumber\\
&=-\frac{2}{\sqrt{\pi}}\lim_{u\rightarrow0}\rmTr_xu\frac{\rmd{\cal H}}{\rmd\ky}
e^{-u^2{\cal H}^2}.
\end{alignat}
Thus, the $\rmTr\partial_1J_1$ term in (\ref{IndFor}) can be integrated as
\begin{alignat}1
\frac{1}{2}\int\rmd^2t\partial_1J_1&
=\int\rmd k_y\frac{\partial}{\partial k_y}\int\rmd x J_1(k_y,x)
\nonumber\\
&=\int\rmd\ky\frac{\rmd\eta(\ky)}{\rmd \ky}
\nonumber\\
&=-\frac{1}{\sqrt{\pi}}\lim_{\mu\rightarrow\infty}\int\rmd\ky
\rmTr_x\frac{1}{\mu}\frac{\rmd{\cal H}}{\rmd\ky}
e^{-{\cal H}^2/\mu^2},
\label{EtaFin}
\end{alignat}
where we have set $u=1/\mu$. Therefore, we see that the leading term of $\rmTr \partial_1J_1$
with respect to $v$ is the continuous part of the spectral asymmetry of ${\cal H}$.
It turns out that this term (\ref{EtaFin}) cancels out the $\tau_3$ term 
(\ref{AnoFin}).

\subsubsection{Calculation of the $\rmTr\partial_2J_2$ term}

Because of the cancellation of the two terms above, 
the remaining $\rmTr\partial_2J_2$ term is solely involved in the index.
Carrying out the integration of the $\partial_2J_2$ term over $t_2=x$ yields
\begin{alignat}1
\int\rmd^2t\partial_2J_2=\int\rmd\ky\left[J_2(\ky,\infty)-J_2(\ky,-\infty)\right] .
\end{alignat}
Therefore, we need to compute the current $J_2$ itself at the spatial infinities.
In the limit of $t_2=x\rightarrow\pm\infty$, we have assumed $m(x)\rightarrow m_\pm=\mbox{const}$.
Then, we can replace ${\cal H}(-i\partial_x,k_y,x)$ by ${\cal H}^\rmB(-i\partial_x,k_y)$, 
\begin{alignat}1
\widetilde{\cal H}&\rightarrow -iv\tau_1\partial_\ky+\tau_2{\cal H}^\rmB(-i\partial_x,k_y).
\end{alignat}
Then, we have
\begin{alignat}1
e^{-i\Omega t}\widetilde{\cal H}e^{i\Omega t}&=
-i\tau_1\left[iv\omega+v\partial_\ky-\tau_3{\cal H}^\rmB(k_x-i\partial_x,k_y)\right]
\nonumber\\
&=-i\tau_1\left[iv\omega+v\partial_\ky-\tau_3{\cal H}^\rmB(k_x,k_y)\right]
+O(\partial_x).
\end{alignat}
Here, it should be noted that $\partial_x$ is irrelevant, since $m$ is constant.
Using these and Appendix \ref{s:CurJ2}, we have
\begin{alignat}1
&J_2(\ky,\pm\infty)
=-i\int\frac{\rmd^2\Omega}{(2\pi)^2}
\rmtr\1_\tau\left(\frac{\partial{\cal H}^\rmB}{\partial k_x}\right)
\frac{1}{iv\omega+v\partial_\ky-\tau_3{\cal H}^\rmB}.
\label{J2DerX}
\end{alignat}
Now let us make the scale transformation $\omega\rightarrow \omega/v$ and 
take the limit $v\rightarrow0$. Then, the series expansion 
with respect to $v$ yields
\begin{alignat}1
J_2(&\ky,\pm\infty)
\nonumber\\
=&-i\int\frac{\rmd^2\Omega}{(2\pi)^2}
\rmtr \1_\tau
\left(\frac{\partial{\cal H}^\rmB}{\partial k_x}\right)
\bigg[\frac{1}{v}\frac{1}{i\omega-\tau_3{\cal H}^\rmB}
-\frac{1}{i\omega-\tau_3{\cal H}^\rmB} \partial_\ky \frac{1}{i\omega-\tau_3{\cal H}^\rmB}
\bigg]+O(v)
\nonumber\\
=&i\int\frac{\rmd^2\Omega}{(2\pi)^2}
\rmtr
\left(\frac{\partial{\cal H}^\rmB}{\partial k_x}\right)
\bigg[\frac{1}{v}\frac{i\omega}{\omega^2+({\cal H}^{\rmB})^2}
+2\frac{1}{i\omega-{\cal H}^\rmB} \partial_\ky \frac{1}{i\omega-{\cal H}^\rmB}
\bigg]+O(v),
\end{alignat} 
where the trace in the $\tau$ space has been taken in the second line.
The first term, being an odd function of $\omega$, vanishes, whereas the second term remains finite.
It can be written using the Green function of the bulk system,
\begin{alignat}1
G(i\omega,k_x,k_y)=\frac{1}{i\omega-{\cal H}^\rmB(k_x,k_y)} ,
\end{alignat}
such that
\begin{alignat}1
J_2
(\ky,\pm\infty)
&=2i
\int\frac{\rmd^2\Omega}{(2\pi)^2}
\rmtr
\left(\frac{\partial{\cal H}^\rmB}{\partial k_x}\right)
\frac{1}{i\omega-{\cal H}^\rmB}
\partial_\ky\frac{1}{i\omega-{\cal H}^\rmB}
\nonumber\\
&=2i
\int\frac{\rmd^2\Omega}{(2\pi)^2}
\rmtr \left(\frac{\partial {\cal H}^\rmB}{\partial k_x}\right)
G\frac{\partial G}{\partial\ky} .
\label{J2CheNum}
\end{alignat}
As we will show in Appendix \ref{s:Che}, this becomes the Chern number $c(m_\pm)$ 
after the integration over $k_y$. Thus, we reach the final expression
\begin{alignat}1
\lim_{M\rightarrow0}
\int_{-\infty}^\infty \rmd k_yJ_2(\ky,\pm\infty)=2c(m_\pm).
\label{J1Ter}
\end{alignat}

\subsection{Result for the index}

Thus far, we have calculated each term in the r.h.s. of the index theorem (\ref{IndFor}).
Equations (\ref{EtaFin}) and (\ref{J1Ter}) allow us to write the index as
\begin{alignat}1
{\rm ind}\,\widetilde{\cal H}=\rmTr\tau_3
-\frac{1}{2}\int\rmd k_y\frac{\rmd\eta}{\rmd k_y}-\left[c(m_+)-c(m_-)\right] ,
\end{alignat}
where the second term in the r.h.s. 
containing the integral represents the continuous part of the 
spectral asymmetry $\eta$. 
This is analogous to the family index theorem derived by Niemi and Semenoff.\cite{NieSem86}
As shown in the calculations above, 
the first and second terms become eqs. (\ref{AnoFin}) and (\ref{EtaFin}), respectively,  
and the index finally becomes 
\begin{alignat}1
-{\rm ind}\,\widetilde{\cal H}&=c(m_+)-c(m_-) .
\label{IndTheFin}
\end{alignat}
It then follows from eq. (\ref{IndEta}) that the spectral flow of the edge states is equal to 
the difference in the Chern numbers,
\begin{alignat}1
\nu=N_+-N_-=c(m_+)-c(m_-).
\end{alignat}
Thus, we establish the bulk-edge correspondence valid not only for the Dirac Hamiltonian but also 
for a wider class of Hamiltonians with higher-order derivative terms.

\section{Summary and Discussions}
\label{s:Sum}

We have explored the bulk-edge correspondence for topological insulators and superconductors with 
broken time reversal symmetry.
We have shown that the spectral flow of the edge states, which are located near a domain wall
separating two topologically distinct bulk states, is identical to the difference in the Chern numbers 
characterizing these two bulk states.
To show this, we have utilized the index theorem developed by Callias, Weinberg, Niemi, and Semenoff,
which is valid for open spaces.
To this end, we have utilized several techniques.
One is the dimensional extension of a one-dimensional Hamiltonian describing the edge states to 
a two-dimensional Hamiltonian, which was often used for index theorems in previous studies.
Another is the introduction of the parameter $v$ in eq. (\ref{ExtHam}) in the dimensional
extension. This allows us to expand the topological index with respect to this parameter.
Since we have assumed Hamiltonians including in general not only the linear kinetic terms but also 
those with higher derivatives, the proof of the correspondence in this paper is valid 
not only for the Dirac Hamiltonian but also for generic nonrelativistic Hamiltonians.   

The generalization of the present approach to the bulk-surface correspondence of three-dimensional 
topological insulators and superconductors, as well as to various topological classes other than
Chern insulators will be interesting.
As to a lattice model, Hatsugai has shown the bulk-edge correspondence for the tight-binding model
on a square lattice in a uniform magnetic field by using the transfer matrix method.
It will be interesting to develop a lattice version of the present method applicable to 
generic lattice models.

\begin{acknowledgements}
This work was supported by a ``Topological Quantum Phenomena'' 
(No. 23103502) Grant-in-Aid for Scientific Research 
on Innovative Areas and Grants-in-Aid for Scientific Research (Nos. 23102714 and 23540406)
from the Ministry of Education, Culture, Sports, Science and Technology
(MEXT) of Japan.
\end{acknowledgements}

\appendix

\section{Detailed calculation of the divergence of the current}
\label{s:DivCur}

In this Appendix, we calculate the r.h.s. of eq. (\ref{RewDivCur}).
It follows from definition (\ref{DefCur}) that the divergence of the current is
\begin{alignat}1
\partial_jJ_j(t)
&=\partial_j^t\lim_{s\rightarrow t}\rmtr\tau_3\Gamma_j
\frac{i}{iM-\widetilde{\cal H}}
\delta^2(t-s)
\nonumber\\
&=\lim_{s\rightarrow t}\rmtr\tau_3\Gamma_j(\partial_j^t+\partial_j^s)
\frac{i}{iM-\widetilde{\cal H}}\delta^2(t-s) ,
\end{alignat}
where the repeated index $j$ should be, of course, summed over.  In what follows, we will compute 
the terms $\partial_1J_1$ and $\partial_2J_2$ separately.
First, $\partial_1J_1$ becomes 
\begin{alignat}1
\partial_1
J_1(t)
=&\lim_{s\rightarrow t}\rmtr\tau_3\Gamma_1
\left[\partial_1^t
\frac{i}{iM-\widetilde{\cal H}}
+
\frac{i}{iM-\widetilde{\cal H}}
\partial_1^s\right]\delta^2(t-s)
\nonumber\\
=&\lim_{s\rightarrow t}\rmtr\tau_3\Gamma_1\left[\partial_1^t
\frac{i}{iM-\widetilde{\cal H}}
-\frac{i}{iM-\widetilde{\cal H}}
\partial_1^t\right]\delta^2(t-s)
\nonumber\\
=&-\lim_{s\rightarrow t}\rmtr\tau_3
\bigg[\gamma_1(-i\partial_1^t)
\frac{1}{iM-\widetilde{\cal H}}
+\frac{1}{iM-\widetilde{\cal H}}
\gamma_1(-i\partial_1^t)\bigg]\delta^2(t-s) .
\end{alignat}
Next, for $\partial_2J_2$,  
we use eq. (\ref{PolIde}),
\begin{alignat}1
\partial_2
J_2(t)
=&-\lim_{s\rightarrow t}\rmtr\tau_3\tau_2
\sum_{n\ge1}\gamma_x^{(n)}\left[(-i\partial_2^t)^n-(i\partial_2^s)^n\right]
\frac{1}{iM-\widetilde{\cal H}}
\delta^2(t-s)
\nonumber\\
=&-\lim_{s\rightarrow t}\rmtr\tau_3\tau_2
\sum_{n\ge1}\gamma_x^{(n)}\bigg[(-i\partial_2^t)^n\frac{1}{iM-\widetilde{\cal H}}
-\frac{1}{iM-\widetilde{\cal H}}(i\partial_2^s)^n\bigg]\delta^2(t-s)
\nonumber\\
=&-\lim_{s\rightarrow t}\rmtr\tau_3
\bigg[\gamma_2(-i\partial_2^t)\frac{1}{iM-\widetilde{\cal H}}
+\frac{1}{iM-\widetilde{\cal H}}\gamma_2(-i\partial_2^t)\bigg]\delta^2(t-s) .
\label{DivCur}
\end{alignat}
Taking these into account, we obtain
\begin{alignat}1
\partial_jJ_j(t)=\lim_{s\rightarrow t}
\rmtr\tau_3
\bigg[(\delta\widetilde{\cal H}-\widetilde{\cal H})\frac{1}{iM-\widetilde{\cal H}}
+\frac{1}{iM-\widetilde{\cal H}}(\delta\widetilde{\cal H}-\widetilde{\cal H})
\bigg]\delta^2(t-s).
\end{alignat}
Note that $\{\delta\widetilde{\cal H},\tau_3\}=0$. Thus, we obtain
\begin{alignat}1
\partial_jJ_j(t)&=-2\lim_{s\rightarrow t}\rmtr\tau_3
\frac{\widetilde{\cal H}}{iM-\widetilde{\cal H}}
\delta^2(t-s).
\end{alignat}


\section{Detailed calculation of  $J_2$
}
\label{s:CurJ2}

In the definition of $J_2$ in eq. (\ref{DefCur}), $\Gamma_2$ in eq. (\ref{DefExtGam})
is not a simple matrix but includes complicated derivatives.
We show in this Appendix that in the plane wave basis, 
it can be written simply as the derivative of ${\cal H}^\rmB$ with respect to $k_x$, as in  eq. (\ref{J2DerX}).
Representing the $\delta$ function as the plane wave, we see that
\begin{alignat}1
J_2(\ky,
\pm\infty)
=&\lim_{s\rightarrow t}
\rmtr\tau_3\Gamma_2\frac{i}{iM-\widetilde{\cal H}}\delta^2(t-s)
\nonumber\\
=&\lim_{s\rightarrow t}\int\frac{\rmd^2\Omega}{(2\pi)^2}
\rmtr\tau_3\Gamma_2
\frac{i}{iM+i\tau_1
\left[v\partial_\ky-\tau_3{\cal H}^\rmB(k_x,\ky)\right]}e^{i\Omega (t-s)}.
\end{alignat}
Using (\ref{PolDef}), we have
\begin{alignat}1
\Gamma_2e^{i\Omega (t-s)}&=e^{i\Omega (t-s)}\tau_2\sum_{n\ge1}\gamma_x^{(n)}p_n(k_x-i\partial_2^t,-k_x-i\partial_2^s)
\nonumber\\& 
=e^{i\Omega (t-s)}\tau_2\sum_{n\ge1}\gamma_x^{(n)}p_n(k_x,-k_x)+O(-i\partial_2) .
\end{alignat}
Here, the derivative terms can be neglected, since ${\cal H}^\rmB$ does not depend on $x$.
Now, note that 
\begin{alignat}1
p_n(k_x,-k_x)=\sum_{0\le i\le n-1}(-)^ik_x^{n-i-1}(-k_x)^i=nk_x^{n-1}.
\end{alignat}
Then, we obtain
\begin{alignat}1
\sum_{n\ge1}\gamma_x^{(n)}p_n(k_x,-k_x)
&=\sum_{n\ge1}\gamma_x^{(n)}nk_x^{n-1}
=\frac{\partial{\cal H}^\rmB}{\partial {k_x}} .
\end{alignat}
Therefore, after taking the $M\rightarrow 0$ limit as well as $s\rightarrow t$, we have
\begin{alignat}1
J_2(\ky,\pm\infty)&=\int\frac{\rmd^2\Omega}{(2\pi)^2}
\rmtr\tau_3\tau_2\left(\frac{\partial{\cal H}^\rmB}{\partial k_x}\right)
\frac{1}{\tau_1\left[iv\omega\tau_1+v\partial_\ky-\tau_3{\cal H}^\rmB\right]} .
\end{alignat}

\section{Calculation of the Chern number}
\label{s:Che}

We show that the r.h.s. of eq. (\ref{J2CheNum}) is simply the Chern number in eq. (\ref{DefCheNum}).
Note that the former equation can be written as
\begin{alignat}1
\frac{1}{2}\int\rmd \ky J_2(\ky)
&=\frac{i}{(2\pi)^2}\int\rmd^3k\rmtr
\left(\frac{\partial{\cal H}^\rmB}{\partial k_x}\right)
G^2\left(\frac{\partial{\cal H}^\rmB}{\partial k_y}\right)G ,
\end{alignat}
where $\rmd^3k\equiv\rmd k_y\rmd\Omega=\rmd^2 k\rmd\omega$.
Inserting a partition of unity,
\begin{alignat}1
\sum_\alpha\psi_\alpha(k)\psi_\alpha^\dagger(k)=1,
\end{alignat}
as well as
\begin{alignat}1
\psi_\beta(\partial_{k_i}{\cal H}^\rmB)\psi_\alpha=(\partial_{k_i}\varepsilon_\alpha)\delta_{\beta\alpha}
-(\varepsilon_\beta-\varepsilon_\alpha)\psi_\beta^\dagger(\partial_{k_i}\psi_\alpha),
\end{alignat}
where $\varepsilon_\alpha(k)$ and $\psi_\alpha(k)$ are respectively the eigenvalue and corresponding eigenfunction
of the bulk Hamiltonian defined in eq. (\ref{BulEigValEqu}),  
we obtain
\begin{alignat}1
\frac{1}{2}\int\rmd 
\ky J_2(\ky)
=&
\frac{i}{(2\pi)^2}\int\rmd^3k\sum_{\alpha,\beta}
\psi_\alpha^\dagger(\partial_{k_x}{\cal H}^\rmB)\psi_\beta
\frac{1}{(i\omega-\varepsilon_\beta)^2}
\psi_\beta^\dagger(\partial_{k_y}{\cal H}^\rmB)\psi_\alpha\frac{1}{i\omega-\varepsilon_\alpha}
\nonumber\\
=&
\frac{-i}{(2\pi)^2}\int\rmd^3k\sum_{\alpha,\beta}
\frac{(\varepsilon_\alpha-\varepsilon_\beta)^2}{(i\omega-\varepsilon_\alpha)(i\omega-\varepsilon_\beta)^2}
\psi_\alpha^\dagger(\partial_{k_x}\psi_\beta)
\psi_\beta^\dagger(\partial_{k_y}\psi_\alpha) .
\end{alignat}
The integration over $\omega$ yields
\begin{alignat}1
&\frac{1}{2}\int\rmd \ky J_2(\ky)
=\frac{i}{2\pi}\int\rmd^2k\sum_{\varepsilon_\alpha<0<\varepsilon_\beta}
\left[
(\partial_{k_x}\psi_\alpha^\dagger)\psi_\beta\psi_\beta^\dagger(\partial_{k_y}\psi_\alpha)-(k_x\leftrightarrow k_y)
\right].
\end{alignat}
It turns out that this is simply the Chern number computed in eq. (\ref{DefCheNum}).




\begin{thebibliography}{9}
\bibitem{footnote2}
Generally, it is expected that for sufficiently large $|k_y|$, $\varepsilon_j(k_y)$ merges into 
the bulk energy continuum in high-energy regions. $\Lambda_{j{\rm U}}$ ($-\Lambda_{j{\rm L}}$)   
may be regarded as the largest (smallest) value 
of $k_y$ at which this occurs. Of course, $\varepsilon_j(k_y)$ is still a continuous function of $k_y$ 
even after merging into the bulk continuum when the system has two open boundaries. 
However, for semi-infinite systems, it may be that
$\varepsilon_j(k_y)$ is not defined for large $|k_y|$ at which the state is absorbed deep into the bulk. 
The argument presented in this paper is also applicable to such situations.
\end{thebibliography}
\end{document}